\documentclass[twocolumn,showpacs,preprintnumbers,amsfonts,amssymb,amsmath]{revtex4}

\usepackage[dvips]{graphicx}
\usepackage{mathrsfs}
\usepackage{amsmath,amssymb}

\usepackage{graphicx}
\usepackage{hhline}
\usepackage{dcolumn}
\usepackage{bm}

\usepackage[dvips]{color}

\begin{document}


\title{Absence of the non-percolating phase for percolation on the non-planar Hanoi network}
\date{\today}

\author{Takehisa Hasegawa}
\email{hasegawa@m.tohoku.ac.jp}
\affiliation{Graduate School of Information Sciences, Tohoku University, 6-3-09, Aramaki-Aza-Aoba, Sendai, 980-8579, Japan}
\author{Tomoaki Nogawa}
\affiliation{Department of Mathematics, Tohoku University, 6-3-09, Aramaki-Aza-Aoba, Sendai, Miyagi 980-8579, Japan}
\begin{abstract}
We investigate bond percolation on the non-planar Hanoi network (HN-NP), which was studied in [Boettcher {\it et al.} Phys. Rev. E 80 (2009) 041115]. 
We calculate the fractal exponent of a subgraph of the HN-NP, which gives a lower bound for the fractal exponent of the original graph. 
This lower bound leads to the conclusion that the original system does not have a non-percolating phase, where only finite size clusters exist, for $p>0$, 
or equivalently, that the system exhibits either the critical phase, where infinitely many infinite clusters exist, or the percolating phase, where a unique giant component exists.
Monte Carlo simulations support our conjecture.
\end{abstract}


\maketitle



\section{introduction}

Percolation is the simplest model exhibiting a phase transition \cite{stauffer1992introduction}. 
Many results for percolation on Euclidean lattices have been reported.
It is well known that bond percolation with open bond probability $p$ on the $d (\ge 2)$-dimensional Euclidean lattice shows a second order transition between 
the non-percolating phase, where only finite size clusters exist, and the percolating phase, where a unique giant component almost surely exists, at a unique critical point $p_c$.
However, this may not be the case for non-Euclidean lattices.

Complex networks have been actively studied in recent years \cite{albert2002statistical,newman2003structure,barrat2008dynamical}.
Among extensive researches carried out on complex networks, 
percolation on various networks has played an important role in clarifying the interplay between network topology and critical phenomena \cite{dorogovtsev2008critical}. 
Percolation on uncorrelated networks (represented by the configuration model \cite{molloy1995critical}) is well described by the local tree approximation; 
there is a phase transition between the non-percolating phase and the percolating phase just as in Euclidean lattice systems, 
but its critical exponents depend crucially on the heterogeneity of the degree distribution of the network \cite{cohen2002percolation}.
On the other hand, several authors \cite{callaway2001randomly,dorogovtsev2001anomalous,zalanyi2003properties,hasegawa2010generating} have reported 
that percolation on networks constructed with certain growth rules exhibits quite a different phase transition from that of uncorrelated networks and Euclidean lattices, 
referred to as an infinite order transition with inverted Berezinskii--Kosterlitz--Thouless (BKT) singularity \cite{hinczewski2006inverted}: 
(i) The singularity of the phase transition is infinitely weak. 
When $p$ lies above the transition point $p_c$, the order parameter $m(p) \equiv \lim_{N \to \infty} s_{\rm max}(N; p)/N$, 
where $s_{\rm max}(N; p)$ is the mean size of the largest cluster over percolation trials in the system with $N$ nodes, obeys 
$m(p) \propto \exp [-{\rm const.}/(\Delta p)^{\beta'}]$, where $\Delta p=p-p_{c}$.
(ii) Below the transition point, the mean number $n_s$ of clusters with size $s$ per node obeys the power law, $n_s \propto s^{-\tau}$.
Furthermore, recent study \cite{boettcher2012ordinary} shows that a hierarchical small-world network exhibits a discontinuous transition instead of an infinite order transition. 

In \cite{hasegawa2010profile,hasegawa2010generating,hasegawa2010critical}, 
it has been found that, for the growing network models the region below $p_c$ corresponds to the critical phase (the intermediate phase), 
which has been observed in nonamenable graphs (NAGs) \cite{benjamini1996percolation,lyons2000phase}. 
NAGs are defined to be transitive graphs with a positive Cheeger constant.
Percolation on NAGs (with one end) exhibits the following three phases depending on the value of $p$: 
the non-percolating phase ($0 \le p<p_{c1}$), the critical phase ($p_{c1}<p<p_{c2}$), where infinitely many infinite clusters exist, 
and the percolating phase ($p_{c2}<p \le 1$).
Here, an infinite cluster is defined to be a cluster whose size is of order $O(N^\alpha)$ ($0< \alpha \le1$). 
It is called a giant component when $\alpha =1$.
In the critical phase, where $0<\alpha<1$, the system is always in a critical state where $n_s$ satisfies a power law \cite{nogawa2009monte}.

\begin{figure}
 \begin{center}
  \includegraphics[width=75mm]{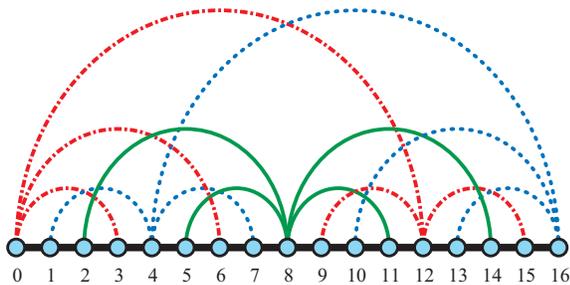}
 \end{center}
 \caption{
(Color online) HN-NP with $L = 4$ generations. The black-thick lines are the backbones, and the red-dashed, green-solid, and blue-dotted lines are the shortcuts of the skeletons $T_a(4)$, $T_b(4)$, and $T_c(4)$, respectively. 
}
 \label{fig-HNNP}
\end{figure}

All previous studies \cite{callaway2001randomly,dorogovtsev2001anomalous,zalanyi2003properties,berker2009critical,hasegawa2010critical,hasegawa2010generating,hasegawa2010profile} 
of percolation on growing networks and hierarchical small-world networks indicate $0=p_{c1} < p_{c2}<1$, except in the following case.
Boettcher {\it et al.} investigated bond percolation on the non-planar Hanoi network (HN-NP) using the renormalization group technique \cite{boettcher2009patchy}. 
They concluded that there are two critical probabilities $p_{c1}$ and $p_{c2}$ between zero and one: $0<p_{c1}<p_{c2}<1$. 

In this paper, we reconsider this model.
We show analytically that the fractal exponent of a subgraph, which is a lower bound for that of the HN-NP, 
takes a non-zero value at all $p (\neq 0)$, indicating that $p_{c1}=0$. 
This means that the system is either in the critical phase or the percolating phase, not in the non-percolating phase, in contrast to the result of \cite{boettcher2009patchy}.
The Monte Carlo simulations support our analytical prediction.


\section{model}

The HN-NP consists of a one-dimensional chain and long-range edges.
The HN-NP with $L (\ge 2)$ generation is constructed as follows \cite{boettcher2009patchy}. 
(i) Consider a chain of $N_L=2^L+1$ nodes. Here, each node $i (=0, 1, 2, \cdots, N_L-1)$ connects to node $i+1$. We call these edges the {\it backbones}.
(ii) For each combination of $i (=0,1,2,\cdots, L-2)$ and $j (=0,1,2,\cdots, 2^{L-i-2}-1)$, nodes $(4j)2^i$ and $(4j+1)2^i$ are connected to $(4j+3)2^i$ and $(4j+4)2^i$, respectively. 
We call these edges the {\it shortcuts}. The schematic of the HN-NP with $L=4$ generations is shown in Fig.\ref{fig-HNNP}.
At generation $L$, the number of backbones is $2^L$ and the number of shortcuts is $2^{L}-2$ (the total number of edges $E_L$ is $E_L=2^{L+1}-2$).
The geometrical properties of the HN-NP are as follows \cite{boettcher2009patchy}:
(i) the degree distribution $p_k$ decays exponentially as $p_{2m+3} \propto 2^{-m}$, 
(ii) the average degree $\langle k \rangle$ is $\langle k \rangle=2E_L/N_L=(2^{L+2} - 4)/(2^L+1) \approx 4$ (for $L \gg 1)$,
(iii) the mean shortest path length $\langle l \rangle$ increases logarithmically with $N_L$ as $\langle l \rangle \propto \log N_L$,
and (iv) the clustering coefficient is zero.

Boettcher {\it et al.} studied bond percolation on the HN-NP with open bond probability $p$ \cite{boettcher2009patchy}. 
In the HN-NP with $L$ generations, they considered the renormalization of four parameters: 
$R_L$ (the probability that three consecutive points $a, b, c$ of a chain are connected), 
$S_L$ (the probability of $bc$ being connected, but not $a$), 
$U_L$ (the probability of $ac$ being connected, but not $b$), 
and $N'_L$  (the probability that there are no connections among $a$, $b$, and $c$).
From the renormalization group flow for $R_L$ they determined the two critical probabilities as $p_{c1}^{\rm BCZ} \approx 0.319445$ and $p_{c2}^{\rm BCZ} \approx 0.381966$.


\begin{figure}[!b]
 \begin{center}
  \includegraphics[width=75mm]{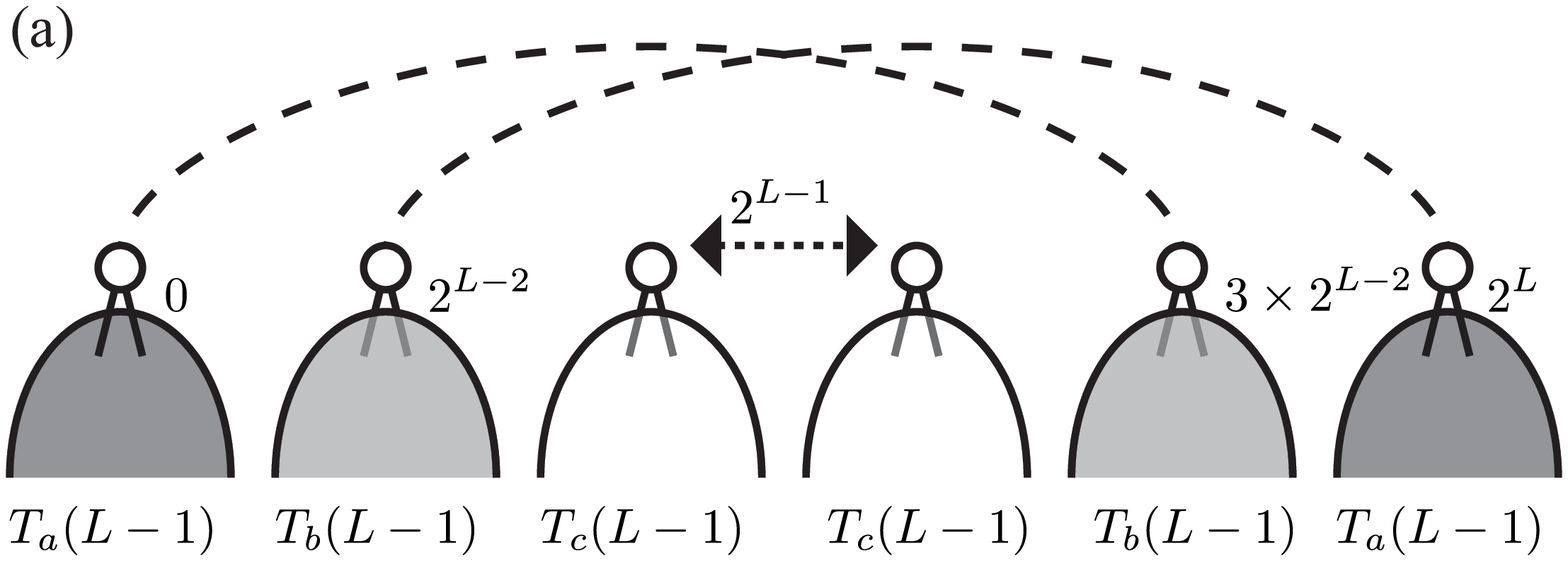}
\\
  \includegraphics[width=75mm]{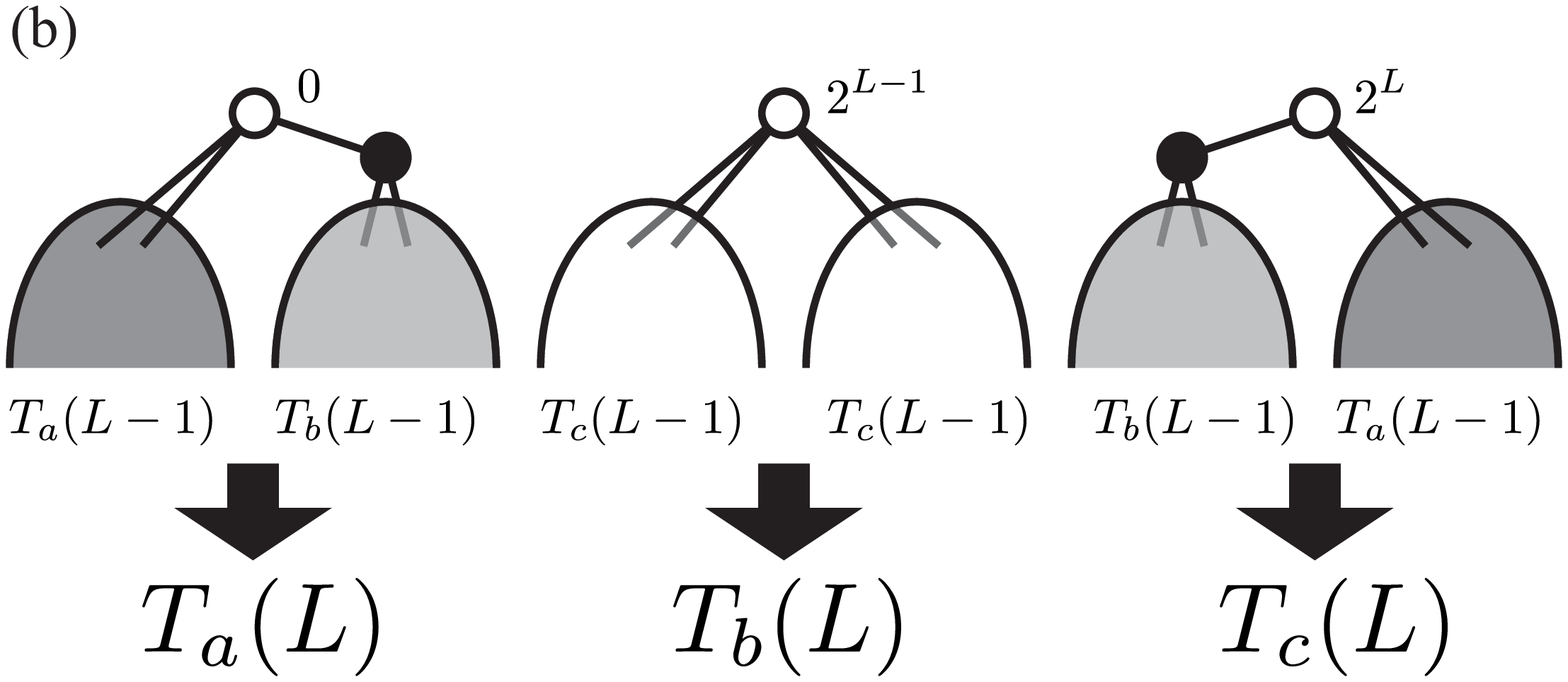}
 \end{center}
 \caption{
Construction of the skeletons $T_a(L)$, $T_b(L)$, and $T_c(L)$. 
Open circles represent the root nodes of the skeletons. 
}
 \label{fig-skeleton}
\end{figure}

\section{analytical calculation for the skeleton of the HN-NP}

The fractal exponent $\psi_{\rm max}(p)$ is useful to determine phase behavior \cite{nogawa2009monte}.
It is defined to be $\psi_{\rm max}(p) = \lim_{N_L \to \infty} \log_{N_L} s_{\rm max}(N_L; p)$. 
A non-percolating phase, a critical phase, and a percolating phase are characterized by $\psi_{\rm max}(p)=0$, $0<\psi_{\rm max}(p)<1$, and $\psi_{\rm max}(p)=1$, respectively.
Unfortunately, it seems difficult to evaluate $s_{\rm max}(N_L; p)$ directly for the HN-NP. 
Instead, we focus on a subgraph of the HN-NP and evaluate its fractal exponent.

We extract a subgraph from the HN-NP with $L$ generations by removing the backbones.
Because the resulting subgraph has no cycles and the number of shortcuts is $2^L-2=N_L-3$, 
this subgraph is composed of three disconnected trees. 
Indeed, nodes $i=0, 2^{L-1}$, and $2^{L}$ belong to the three different trees.
We call these {\it the root nodes}.
Here the graphs isomorphic to these trees and having root nodes $i=0, 2^{L-1}$, and $2^{L}$ will be called the {\it skeletons} $T_a(L)$, $T_b(L)$, and $T_c(L)$, respectively. 
Clearly, $T_a(L)$ and $T_c(L)$ are also isomorphic to each other, $T_a(L) \simeq T_c(L)$.
At $L=2$, $T_a(2)$ is composed of nodes $0$ and 3 and the edge between them, $T_b(2)$ is one isolated node $i=2$, 
and $T_c(2)$ is composed of nodes $1$ and 4 and the edge between them.
The skeletons $T_a(L)$, $T_b(L)$, and $T_c(L)$ for arbitrary $L$ are given recursively as follows.
First, we consider two subgraphs of the sets of nodes $\{0, 1, \cdots, 2^{L-1}\}$ and $\{2^{L-1}, 2^{L-1}+1, \cdots, 2^{L}\}$ after removing the backbones from the HN-NP with $L$ generations.
By symmetry, both subgraphs consist of the skeletons $T_a(L-1)$, $T_b(L-1)$, and $T_c(L-1)$ (Fig.\ref{fig-skeleton}(a)). 
Here the root nodes of the latter subgraph are $i=2^{L-1}$ (for $T_c(L-1)$), $3 \times 2^{L-2}$ (for $T_b(L-1)$), and $2^{L}$ (for $T_a(L-1)$).
Note that the skeletons $T_a(L)$, $T_b(L)$, and $T_c(L)$ are given 
by adding the two long-range edges $\{0, 3 \times 2^{L-2}\}$ and $\{2^{L-2}, 2^{L}\}$, and taking into account the connection of node $2^{L-1}$ (Fig.\ref{fig-skeleton}(b)), 
we have
\begin{eqnarray}
T_a(L)&\simeq& T_c(L) \simeq R_1(T_a(L-1), T_b(L-1)), \label{recrelA} \\
T_b(L)&\simeq&R_2(T_c(L-1), T_c(L-1)), \label{recrelB}
\end{eqnarray}
where the operation $R_1(x,y)$ adds the edge between root nodes of the skeletons $x$ and $y$, and $R_2(x,y)$ merges two root nodes of the skeletons $x$ and $y$ into one.

We now calculate the mean size of a cluster including the root node (the root cluster size) for each skeleton.
We denote the root cluster sizes of $T_a(L) (\simeq T_c(L))$ and $T_b(L)$ by $s_a(L)$ and $s_b(L)$, respectively.
Because of the recursive structure (\ref{recrelA}, \ref{recrelB}) of the skeletons, the root cluster sizes $s_a(L)$ and $s_b(L)$ also satisfy recursive relations:
\begin{eqnarray}
s_a(L+1)&=&s_a(L)+p s_b(L), \\
s_b(L+1)&=&2 s_a(L) -1,
\end{eqnarray}
where the initial conditions are $s_a(2)=1+p$ and $s_b(2)=1$.
Then, we find that 
\begin{eqnarray}
s_a(L)&=&
\frac{1}{2}+\frac{\left(1+\sqrt{1+8 p}\right)^{L+1}-\left(1-\sqrt{1+8 p}\right)^{L+1}}{2^{L+2}\sqrt{1+8 p}}, \\
s_b(L)&=&1+\frac{\left(1+\sqrt{1+8 p}\right)^{L}-\left(1-\sqrt{1+8 p}\right)^{L}}{2^{L}\sqrt{1+8 p}}.
\end{eqnarray}
For $L \gg 1$, we obtain $s_{a}(L) \propto {N_L}^{\psi_{\rm root}^{\rm skeleton}(p)}$, where
\begin{equation}
\psi_{\rm root}^{\rm skeleton}(p) = \log_2(1+\sqrt{1+8p})-1. \label{psiAnalytics}
\end{equation}
We expect $\psi_{\rm root}^{\rm skeleton}(p)=\psi_{\rm max}^{\rm skeleton}(p)$ because the roots are hubs.
In fact, we performed Monte Carlo simulations for the bond percolation on the skeletons. 
Our numerical result of $\psi_{\rm max}^{\rm skeleton}(p)$ shows a good correspondence with Eq.(\ref{psiAnalytics}) except near $p=0$ (not shown).
According to Eq.(\ref{psiAnalytics}), $\psi_{\rm root}^{\rm skeleton}(p)$ increases continuously from $\psi_{\rm root}^{\rm skeleton}(0)=0$ to $\psi_{\rm root}^{\rm skeleton}(1)=1$.
This means that the subsystem consisting only of the shortcuts is in the critical phase for all $p (\neq 0,1)$, like the growing random tree \cite{hasegawa2010critical}.
Because the HN-NP is obtained by adding the backbones to the skeletons, the clusters in the skeletons become larger. 
Therefore, the entire system permits a critical phase even for infinitesimal $p$, i.e., the non-percolating phase does not exist except at $p=0$.

\begin{figure}[!b]
 \begin{center}
  \includegraphics[width=75mm]{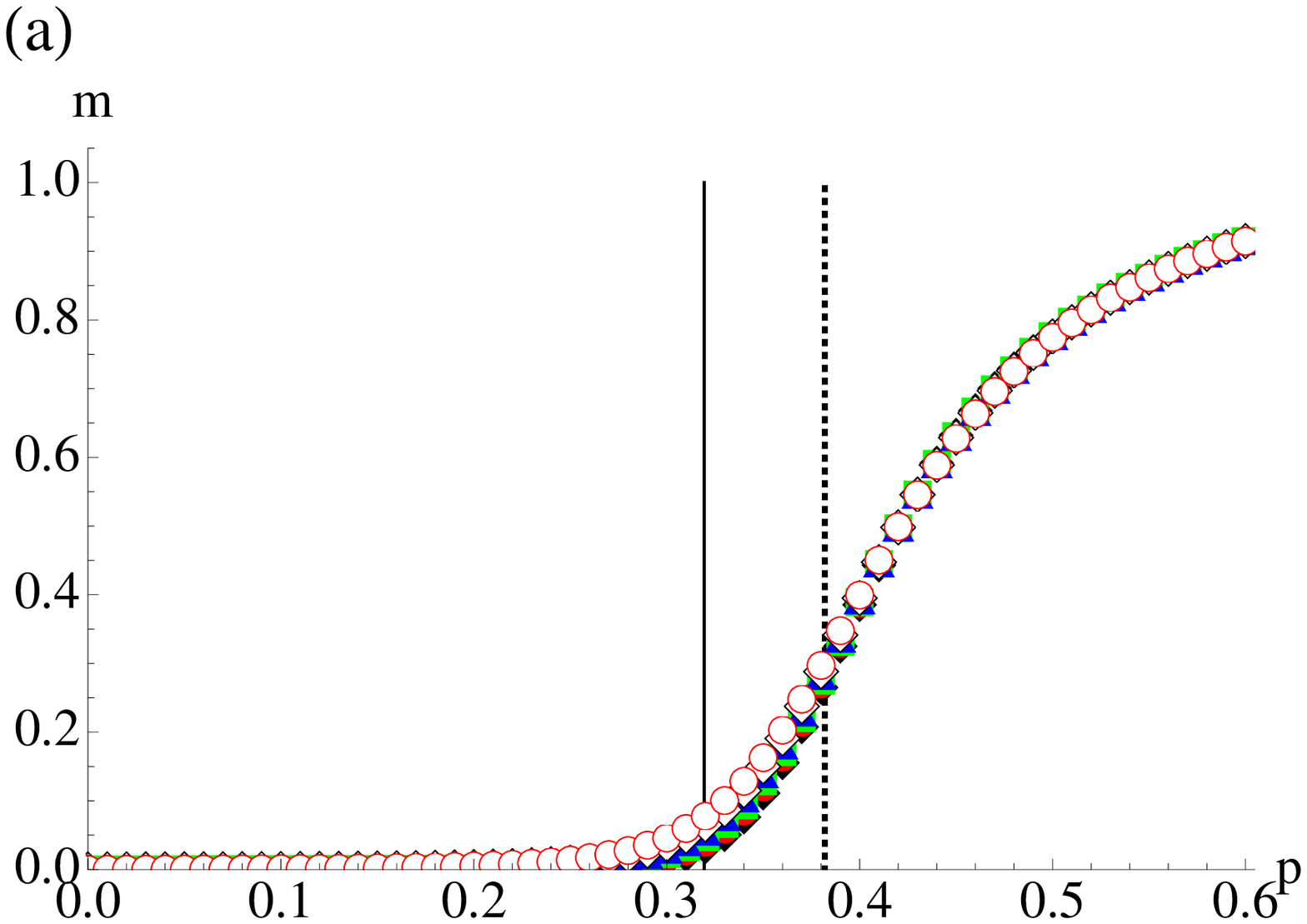}
  \includegraphics[width=75mm]{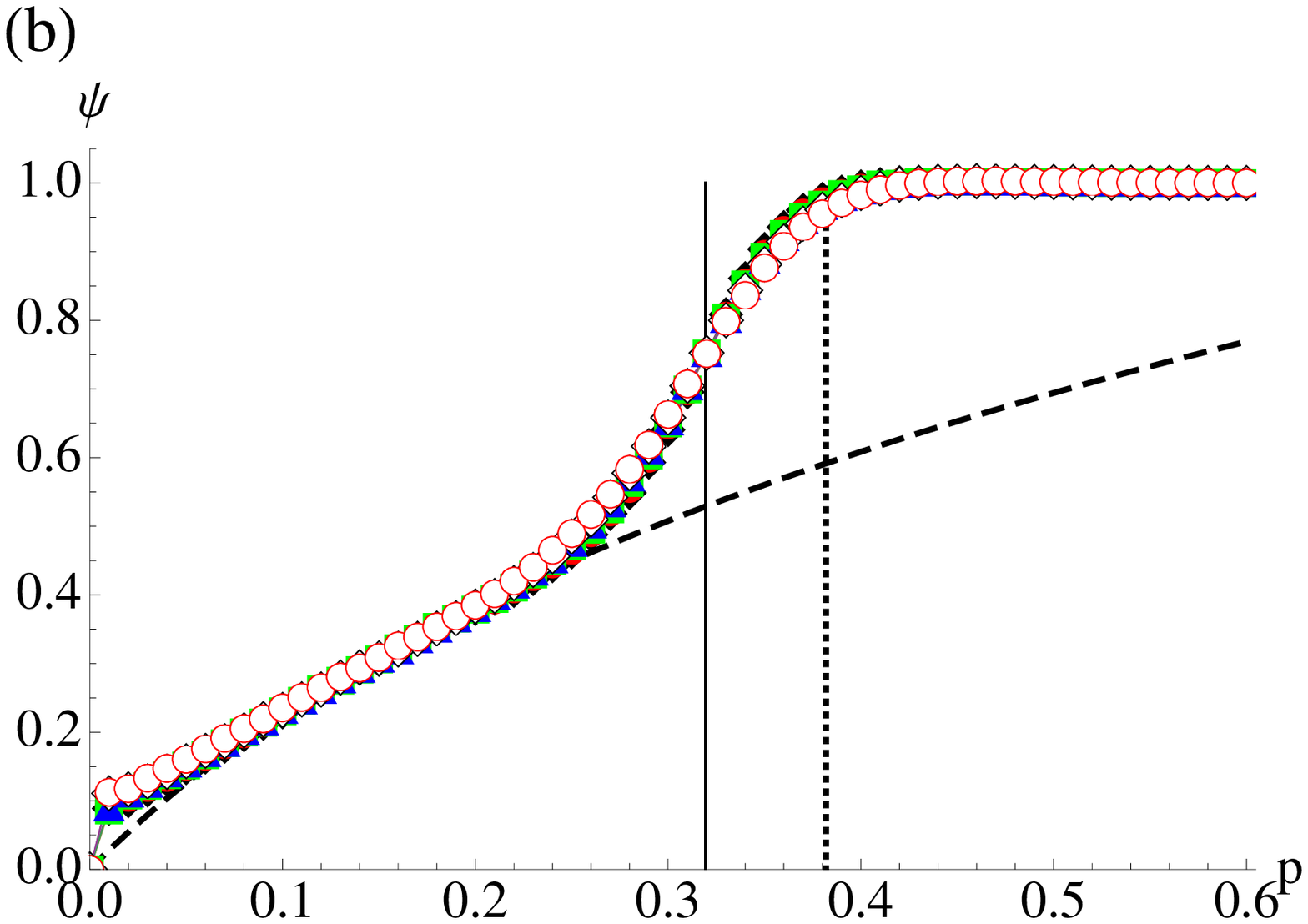}
 \end{center}
 \caption{
(Color online)
 (a) Order parameter $m(N_L; p)=s_{\rm max}(N_L; p)/N_L$ and (b) fractal exponent $\psi(N_L; p)$.
The numbers of generations $L$ are 19 (black-diamond), 18 (red-circle), 17 (green-square), 16 (blue-triangle), 15 (open-diamond), and 14 (open-circle). 
The vertical solid line and dashed line indicate $p_{c1}^{\rm BCZ}$ and $p_{c2}^{\rm BCZ}$ respectively.
In (b), the fractal exponent $\psi_{\rm root}^{\rm skeleton}(p)$ of the skeleton given by Eq.(\ref{psiAnalytics}) is shown by the thick-dashed line.
}
 \label{fig-psi}
\end{figure}

\begin{figure}[!b]
 \begin{center}
  \includegraphics[width=75mm]{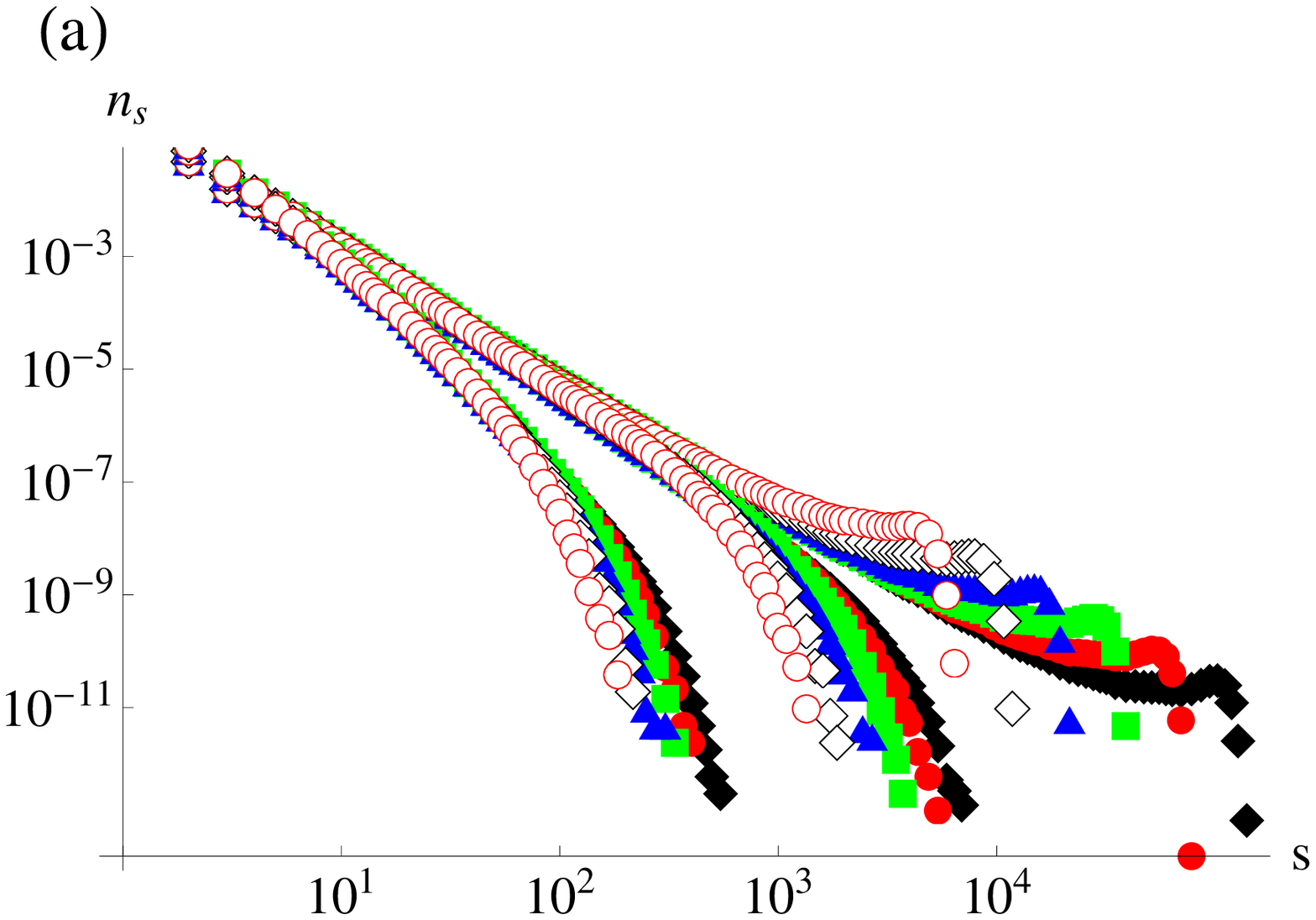}
  \includegraphics[width=75mm]{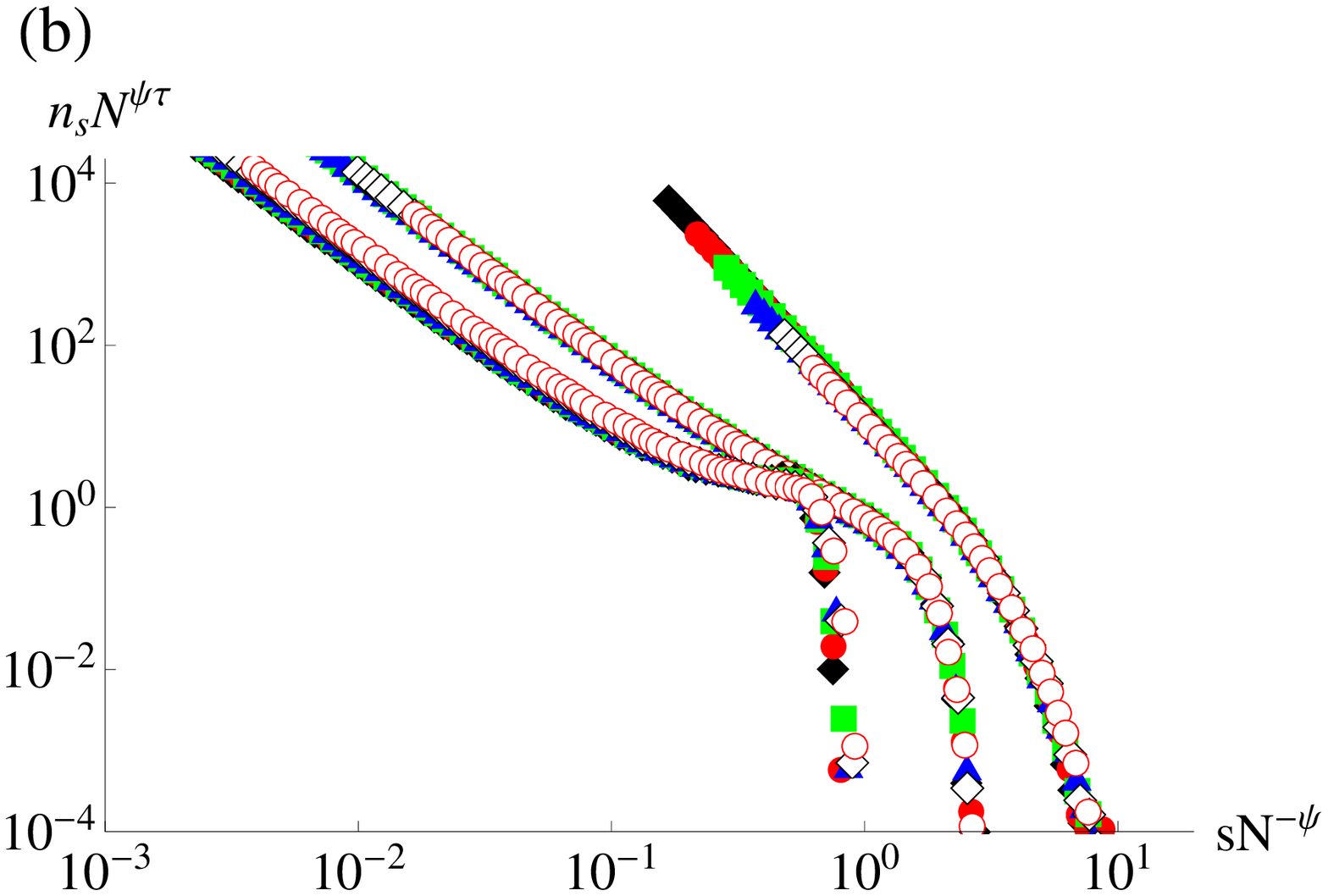}
 \end{center}
 \caption{
(Color online)
(a) Cluster size distribution $n_s(N_L; p)$ at $p=0.16$, $0.26 (<p_{c1}^{\rm BCZ})$, and $0.36 (>p_{c1}^{\rm BCZ})$, from left to right. 
(b) Scaling result for $n_s(N_L; p)$ at $p=0.20 (<p_{c1}^{\rm BCZ})$, 
$0.32$, and $0.35 (>p_{c1}^{\rm BCZ})$, from right to left. 
The numbers of generations $L$ are 19 (black-diamond), 18 (red-circle), 17 (green-square), 16 (blue-triangle), 15 (open-diamond), and 14 (open-circle). 
}
 \label{fig-ns}
\end{figure}


\section{numerical check}

In the previous section, we evaluated the root cluster size of the skeleton to show that its fractal exponent $\psi_{\rm root}^{\rm skeleton}(p)$ takes a non-zero value for all $p >0$. 
Because the skeleton is just a subgraph of the HN-NP, 
$\psi_{\rm root}^{\rm skeleton}(p)$ is a lower bound for the fractal exponent of the largest cluster of the HN-NP $\psi_{\rm max}^{\rm HN-NP}(p)$, 
i.e., $\psi_{\rm root}^{\rm skeleton}(p) \approx \psi_{\rm max}^{\rm skeleton}(p) \le \psi_{\rm max}^{\rm HN-NP}(p)$. 
For bond percolation on the HN-NP, $\psi_{\rm max}^{\rm HN-NP}(p)>0$ when $p>0$, implying that $p_{c1}=0$.
To check our prediction, we performed Monte Carlo simulations of bond percolation on the HN-NP.
The number of generations is $L=13, 14, \cdots, 20$, and the number of percolation trials is 100000 for each $p$.

Figures \ref{fig-psi}(a) and (b) show the results for the order parameter $m(N_L; p)$ and the fractal exponent of the largest cluster $\psi_{\rm max}^{\rm HN-NP}(N_L; p)$, respectively.
Here the fractal exponent $\psi_{\rm max}^{\rm HN-NP}(N_L; p)$ at a finite generation $L$ is evaluated as 
\begin{equation}
\psi_{\rm max}^{\rm HN-NP}(N_L; p) \approx \frac{\log s_{\rm max}(N_{L+1}; p) - \log s_{\rm max}(N_{L-1}; p)}{\log N_{L+1}- \log N_{L-1}}.
\end{equation}
We also plot the fractal exponent $\psi_{\rm root}^{\rm skeleton}(p)$ of the skeleton (Eq.(\ref{psiAnalytics}), shown as the thick-dashed line) 
and $p_{c1}^{\rm BCZ}$ and $p_{c2}^{\rm BCZ}$ (shown as vertical lines) in Fig.\ref{fig-psi}.

From Fig.\ref{fig-psi}(b), we see that $\psi_{\rm root}^{\rm skeleton}(p)$ is actually the lower bound of $\psi_{\rm max}^{\rm HN-NP}(p)$, implying that $p_{c1}=0$.
In particular, $\psi_{\rm max}^{\rm HN-NP}(N_L; p)$ coincides with $\psi_{\rm root}^{\rm skeleton}(p)$ for $p \lesssim 0.26$ (except near $p=0$, where finite size effects are not negligible). 
For $p \gtrsim 0.26$, $\psi_{\rm max}^{\rm HN-NP}(N_L; p)$ is considerably greater than $\psi_{\rm root}^{\rm skeleton}(p)$, and reaches unity at $p=p_{c2}^{\rm BCZ}$.
At a glance, in the large size limit, $\psi_{\rm max}^{\rm HN-NP}(N_L; p)$ seems to change continuously with $p<p_{c2}^{\rm BCZ}$.
However, we speculate that in the large size limit $\psi_{\rm max}^{\rm HN-NP}(N_L; p)$ (i) coincides with $\psi_{\rm root}^{\rm skeleton}(p)$ in the entire region below $p_{c1}^{\rm BCZ}$,  
(ii) jumps to a higher value at $p=p_{c1}^{\rm BCZ}$, and (iii) increases monotonically up to unity for $p_{c1}^{\rm BCZ}<p \le p_{c2}^{\rm BCZ}$.
The coincidence between $\psi_{\rm max}^{\rm HN-NP}(p)$ and $\psi_{\rm root}^{\rm skeleton}(p)$ for $p<p_{c1}^{\rm BCZ}$ means 
that the partial ordering (in the sense that the largest cluster is $O(N^\alpha)$ with $\alpha <1$) in this region is essentially governed by the shortcuts.
Because Boettcher {\it et al.} \cite{boettcher2009patchy} considered renormalization of the connecting probability of consecutive points of the backbones, 
we would expect their first critical probability $p_{c1}^{\rm BCZ}$ to be the probability above which the backbones become relevant.
Thus, we expect that there is a transition between critical phases, in the sense that the fractal exponent jumps, implying a qualitative change in the criticality, while it is very difficult to judge whether such a transition exists or not by finite size simulations.
Such a jump in the fractal exponent has already been observed in site-bond percolation on the decorated (2,2)-flower \cite{hasegawa2012phase}.
In addition, our numerical result shows that $\psi_{\rm max}^{\rm HN-NP}(p)$ reaches unity smoothly at $p_{c2}^{\rm BCZ}$.
This indicates that the phase transition to the percolating phase is discontinuous, similarly as in \cite{boettcher2012ordinary}.

Finally, we discuss the cluster size distribution function, $n_s(p)$, below $p_{c2}^{\rm BCZ}$.
Figure \ref{fig-ns}(a) shows $n_s(p)$ for several values of $p$ with $0<p<p_{c2}^{\rm BCZ}$. 
In the critical phase, we expect a power law for $n_s(p)$: 
\begin{equation}
n_s(p) \propto s^{-\tau(p)}, \label{eq-ns}
\end{equation}
where 
\begin{equation}
\tau(p)=1+\psi_{\rm max}(p)^{-1}, \label{eq-tau}
\end{equation} 
and a corresponding scaling form: 
\begin{equation}
n_s(N_L; p)=N_L^{-\psi_{\rm max}(p)\tau(p)} f(sN_L^{-\psi_{\rm max}(p)}), \label{eq-nsscaling}
\end{equation}
where the scaling function $f(\cdot)$ behaves as 
\begin{equation}
f(x)\sim
\begin{cases}
\text{rapidly decaying func.} & \text{for $x\gg1$,}\\
x^{-\tau(p)} & \text{for $x\ll 1$.}
\end{cases}
\end{equation}
We tested this scaling for $0<p \lesssim 0.26$ and $p_{c1}^{\rm BCZ}<p<p_{c2}^{\rm BCZ}$ and obtained excellent collapses (Fig.\ref{fig-ns}(b)). 
We would also expect that $n_s$ to be fat-tailed for $0.26 \lesssim p<p_{c1}^{\rm BCZ}$ 
because $n_s(p)$ for the skeletons perfectly obeys Eqs.(\ref{eq-ns}) and (\ref{eq-tau}) via Eq.(\ref{eq-nsscaling}) for $0<p<1$ (not shown), 
and $n_s$ is broader when we add the backbones to the skeletons, i.e., for the original HN-NP.


\section{summary}

In this paper, we have studied bond percolation on the HN-NP.
Our results give the two critical probabilities as $p_{c1}=0(<p_{c1}^{\rm BCZ})$ and $p_{c2}=p_{c2}^{\rm BCZ}$, implying that the system has only a critical phase and a percolating phase, 
and does not have a non-percolating phase for $p>0$.
As far as we know, all complex network models with a critical phase have only the critical phase and the percolating phase (\cite{callaway2001randomly,dorogovtsev2001anomalous,zalanyi2003properties,berker2009critical,hasegawa2010critical,hasegawa2010generating,hasegawa2010profile,boettcher2012ordinary} for percolation 
and \cite{bauer2005phase,khajeh2007berezinskii,berker2009critical,boettcher2011fixed,boettcher2011renormalization,nogawa2012generalized,nogawa2012criticality,hinczewski2006inverted} for spin systems).
It will be challenging to clarify the origin of such universal behavior.


\section*{Acknowledgments}
TH acknowledges the support through
Grant-in-Aid for Young Scientists (B) (No. 24740054) from MEXT, Japan.

\end{document}